\documentclass{elsart}
\usepackage{graphicx}
\usepackage{amssymb}
\begin{document}
\begin{frontmatter}
\title{Universality of electron distributions in high-energy air showers - 
description of Cherenkov light production}
%
%
\vspace{-0.3cm}
\author{F. Nerling$^{a,\dag}$, J. Bl\"umer$^{a,b}$, R. Engel$^{a}$, M. Risse$^{a}$}

\address{$^{a}$Forschungszentrum Karlsruhe, Institut f\"ur Kernphysik
\\
$^{b}$Universit\"at Karlsruhe, Institut f\"ur Experimentelle Kernphysik
\\
\vspace{0.5cm}
{\rm $^{\dag}$Corresponding author, Email: frank.nerling@ik.fzk.de}}

\vspace{-0.3cm}
\begin{abstract}
The shower simulation code CORSIKA has been used to investigate the electron energy and angular distributions in 
high-energy showers. 
Based on the universality of both distributions, we develop an analytical description of Cherenkov light emission 
in extensive air showers, which provides the total number and angular distribution of photons.
The parameterisation can be used e.g. to calculate the contribution of direct and
scattered Cherenkov light to shower profiles measured with the air fluorescence technique.
\end{abstract}
\vspace{-0.3cm}
\end{frontmatter}

%
%
\section{Introduction}
\label{INTRO}
The measurement of ultrahigh-energy air showers above $10^{17}$ eV is one 
of the main tools to understand the nature, origin and propagation of cosmic rays at highest energies. 
Large-scale experiments, like the Pierre Auger Observatory \cite{abraham 2004}, AGASA \cite{agasa}, 
HiRes \cite{hires}, and Telescope Array \cite{ta} focus on the precise determination of the energy 
spectrum, mass composition and arrival direction distribution of ultrahigh-energy cosmic rays (UHECR). 

The calorimetric measurement of the longitudinal shower profile with the fluorescence technique is one of the most 
direct methods to determine the primary energy \cite{baltrusaitis 1985} and also can be used to infer the primary 
particle type. 
The charged shower particles induce nitrogen fluorescence light, 
which is emitted isotropically mainly in the near UV-range of 300-400 nm \cite{bunner 1967}. 
Assuming the fluorescence yield is proportional to the energy deposit by a shower, as 
indicated by measurements \cite{kakimoto 1996,nagano 2003}, 
observing the emitted fluorescence light gives a measure of the local ionisation energy deposit 
and hence the shower size. 

However, charged particles in extensive air showers (EAS) are mostly relativistic and Cherenkov light is 
produced as well.
In contrast to the fluorescence light, Cherenkov photons are emitted mostly 
in the forward direction. 
Depending on the observation angle with respect to the shower axis, the intensity of 
produced Cherenkov light contributes a non-negligible signal in 
fluorescence detectors \cite{baltrusaitis 1985,galbraith 1953,nerling 2004,perrone 2003}, 
which can be of same order as the fluorescence light signal itself \cite{nerling 2004,perrone 2003,nerling 2005}.
Thus, an efficient correction for or an explicit consideration of the Cherenkov contribution to the 
measured signal is needed for the determination of the primary particle properties. 

In the pioneering Fly's Eye experiment, Cherenkov light was estimated \cite{baltrusaitis 1985} based 
on Hillas' simulations of 100\,GeV photon showers \cite{hillas 1982}. 
Assuming universality of electron distributions in EAS, the parameterisation in shower age developed by Hillas was 
applied to showers of ultrahigh-energy \cite{baltrusaitis 1985,hillas 1982_b,patterson 1983}. 
More recent simulation studies at higher energies and with hadron
primaries confirm such an universality of electron energy spectra \cite{giller 2004,nerling 2003}.

During the last decades substantial progress has been made in the simulation of very high energy EAS. New simulation
packages such as CORSIKA \cite{heck 1998} and modern high-energy interaction models allow a more reliable prediction
of particle distributions in showers. At the same time the precision of UHECR detectors has increased dramatically,
requiring a treatment of Cherenkov light production as accurate as possible.

In this work we perform a systematic study of shower particle distributions relevant to Cherenkov light
calculation. A complete model for Cherenkov light calculation is developed and compared to predictions from 
CORSIKA employing QGSJET\,01 \cite{kalmykov 1997}. 
In this approach, the total Cherenkov light can be calculated as function either of the shower 
size profile or the energy deposit profile. 
Parameterisations of electron energy spectra and Cherenkov photon angular distributions in high-energy 
showers are developed, which serve as a basis for calculations of radiation 
emitted by air showers and may also be useful for other purposes. 
A direct application is the prediction of Cherenkov radiation as needed for 
reconstruction of longitudinal shower size profiles measured by 
experiments based on the fluorescence technique.

The outline of the paper is as follows. 
In Sec.\,\ref{sec.CHprod}, an analytical expression is derived 
for calculating Cherenkov light production in EAS based on electrons\footnote{Throughout this work, the notation `electrons' include 
$e^{+}$ and $e^{-}$.} only. 
In Sec.\,\ref{sec.ED}, electron energy spectra are shown to be universal, parameterised in terms of a 
phenomenological definition of the shower age parameter and applied for calculating the longitudinal 
Cherenkov light profile; the results are compared to other 
approaches and to a detailed CORSIKA simulation. 
Angular distributions of electrons and Cherenkov photons are investigated in Sec.\,\ref{sec.AD}. The 
former are shown to be universal and the latter are parameterised in dependence of altitude and 
shower age. 
Finally, the concept of ionisation energy deposit in simulation and reconstruction of fluorescence 
light profiles is discussed in Sec.\,\ref{sec.Edep}. 
Part of the results presented here have already been shown in \cite{nerling 2005,nerling 2003,nerling 2005_a}.

%
%
\vspace{-0.3cm}
\section{Calculation of Cherenkov light production \label{sec.CHprod}}
The total number of Cherenkov photons ${\rm d}N_{\gamma}$ produced in a shower per interval of slant depth d$X$ and 
angle d$\theta$ with respect to the shower axis is given by
\begin{equation}
\label{eq.CHprod_allg}
\frac{{\rm d}N_{\gamma}}{{\rm d}X{\rm d}\theta}(X,\theta,h)~=~\sum_{i={\rm e,\mu},...}
A_{\rm \gamma}^{{i}}(X,\theta,h)~\int_{E_{\rm thr}^{i}}^{\infty}~\frac{{\rm d}N_{i}}{{\rm d}E}(X)~~y_{\gamma}^{{i}}(h,E)~~{\rm d} E~, 
\end{equation}
where ${\rm d}N_{i}/{\rm d}E\,(X)$ are the normalised differential energy spectra of corresponding charged
particles of type $i$ (= electrons, muons, ...) at depth $X$, 
$E_{{\rm thr}}^{i}(h)$ the height-dependent Cherenkov energy thresholds, 
and $y_{\gamma}^{i}(h,E)$ the corresponding Cherenkov photon yields. 
$A_{\rm \gamma}^{i}(X,\theta,h)$ are the normalised angular distributions of produced Cherenkov 
photons with respect to the shower axis. 
The photons are assumed to be produced at the shower axis, 
which is a good approximation for fluorescence technique applications 
since most charged particles are moving at a lateral distance of less than 60\,m. 

The amount of so-called {\it scattered Cherenkov light} reaching a detector depends mostly on the 
total number of photons produced since the dominant Rayleigh scattering process is nearly isotropic. 
Therefore, the integral $\frac{{\rm d}N_{\gamma}}{{\rm d}X}(X,h)$, corresponding to integration 
of Eq.\,(\ref{eq.CHprod_allg}) over all 
angles $\theta$, is the most important quantity and typically used to calculate the scattered Cherenkov 
light contribution in fluorescence measurements. 
The so-called {\it direct Cherenkov light} contribution, i.e. Cherenkov photons hitting directly 
the detector without scattering in the atmosphere, is determined by both the total Cherenkov light 
intensity and the angular distributions $A_{\rm \gamma}^{i}(X,\theta,h)$. 

The number of Cherenkov photons produced by a charged particle of total energy $E$ and charge $Z$ in a wavelength interval 
between $\lambda_1$ and $\lambda_2$ is given by 
\begin{equation}
\label{eq.yield}
y_{\gamma} := \frac{{\rm d}N_{\gamma}^{(1)}}{{\rm d}X} (h,E)= \frac {2\pi \alpha  Z^2}{\rho(h)} 
\int_{\lambda_{1}}^{\lambda_{2}} \left(1-\frac{1}{n^{2}(h,\lambda)\,\beta^{2}}\right)\,\frac{\d\lambda}{\lambda^{2}}~~,
\end{equation}
where $\alpha$ is the fine-structure constant, $\beta =v/c$, $\rho$ the air density at height $h$, 
and $n$ the refractive index. 
Measuring d$X$ along the shower trajectory we introduce the approximation that all particle trajectories 
are parallel to the shower axis, resulting in an under-estimation of produced Cherenkov photons of
order one percent, cf. Sec.\,\ref{sec.CHtotnum}. 
The wavelength interval is usually about 300-400\,nm for fluorescence telescopes \cite{abraham 2004,baltrusaitis 1985}.
Since in air $n \approx 1$ and dispersion is negligible the integrand of (\ref{eq.yield}) can be approximated as
\begin{equation}
 1-(\beta n)^{-2} = 1- \left(1-\frac{m^{2}{\rm c}^{4}}{E^{2}}\right)^{-1} (1+\delta)^{-2}\nonumber 
 \approx 2\delta - \frac{m^{2}c^{4}}{E^{2}}~,  
\end{equation}
\noindent
where $\delta(h)= n(h) -1$, and $m$ the charged particle mass. The energy threshold 
condition for Cherenkov radiation in air reads  
\begin{equation}
\label{eq.Ethr}
E_{{\rm thr}}(h) = m{\rm c}^{2}/\sqrt{2\delta(h)}~,
\end{equation}
giving 21\,MeV for electrons and 4.4\,GeV for muons at sea level. These thresholds increase to 37\,MeV and
7.6\,GeV respectively at an altitude of 10\,km (US standard atmosphere).

Particles heavier than electrons contribute less than 2\,\% to the shower size around the shower maximum even for heavy primaries and their energy
threshold for Cherenkov light production is very high. 
Therefore, to a good approximation, practically all charged particles can assumed to be electrons in the following 
considerations. Expression (\ref{eq.CHprod_allg}) simplifies to the following ansatz
\begin{equation}
\label{eq.CHprod}
\frac{{\rm d}N_{\gamma}}{{\rm d}X{\rm d}\theta}(X,\theta,h) = A_{\rm \gamma}(X,\theta,h)\cdot N(X)
\int_{\ln E_{{\rm thr}}}y_{\gamma}(h,E)~f_{\rm e}~(X,E)~{\rm d}\ln E ~.
\end{equation}
Here, $N(X)$ is the charged particle number as function of depth $X$, and $E_{\rm thr}$ the local 
Cherenkov energy threshold for electrons. For a given shower geometry, $h=h(X)$ follows from the 
atmospheric model assumed, $f_{\rm e}(X,E)$ is the normalised differential electron energy 
spectrum at depth $X$ 
\begin{equation}
f_{\rm e}(X,E) = \frac{1}{N_{\rm e}\,(X)}\,\frac{{\rm d}N_{\rm e}}{{\rm d}\ln E}\,(X,E)~,
\end{equation} 
and $A_{\rm \gamma}(X,\theta,h)$ is the normalised angular distribution of all Cherenkov photons 
produced. 

We show in Sec.\,\ref{sec.ED} and \ref{sec.Edep} that the ansatz (\ref{eq.CHprod}) 
describes well the Monte Carlo results used as reference.

%
%
\subsection{Monte Carlo simulations}
\begin{figure}[t]
\begin{center}
\includegraphics[clip,bb= 3 21 564 527,width=0.65\linewidth]{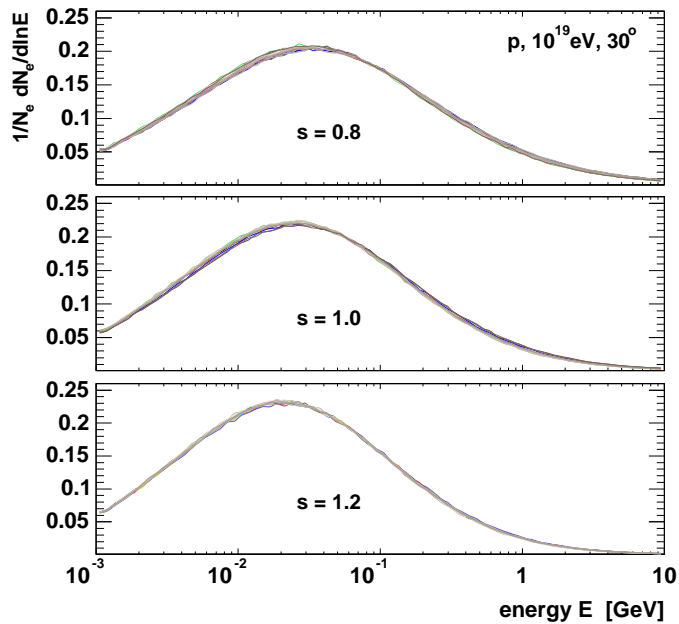}
\end{center}
\caption{Electron energy spectra obtained with CORSIKA for 3 different fixed shower ages. 
Shown are 15 individual proton showers of $10^{19}$\,eV.
\newline~~}
\label{fig.ED_in_s}
\end{figure}
The cosmic ray simulation code CORSIKA \cite{heck 1998} version 6.137 with 
the hadronic interaction models QGSJET\,01 \cite{kalmykov 1997} for high-energy interactions and GHEISHA 2002 
\cite{fesefeldt 1985,cassell 2002} for low-energy interactions has been used to study the development
of high-energy showers. Energy spectra of electrons and angular distributions of electrons and Cherenkov photons 
were obtained for individual showers and different combinations of primary energy, zenith angle and primary 
particle type. The US standard atmosphere \cite{US-StdA,knapp 1993} has been used as atmospheric model. 
Also the longitudinal shower size profile and produced Cherenkov photon profile, 
which are needed to test the analytical model proposed in this paper, are calculated in the simulations. 
In CORSIKA calculations, the wavelength dependence of the
refractive index $n(\lambda)$ is neglected. This simplification is a good approximation as has been shown by a 
comparison \cite{perrone 2003} to GEANT\,3.21 \cite{geant} simulations taking into account the wavelength dependence.
To reduce CPU-time, optimum thinning \cite{kobal 2001,risse 2001} of $10^{-6}$ has been applied unless otherwise noted. 
The UPWARD option has been enabled to follow the upward going electromagnetic 
particles\footnote{In CORSIKA, particles are counted at horizontal layers, which can be
specified by the user. By default, particles with tracks at angles larger than 
$90^{\circ}$ to the vertical are not followed. By applying the UPWARD option, 
these particles are treated in the electromagnetic shower component.}. 
Low-energy thresholds of 100\,MeV for hadrons and muons, and of 1\,MeV for electrons 
and gamma-rays have been used, unless otherwise noted. 

%
%
\section{Electron energy spectra \label{sec.ED}}
\begin{figure}[t]
\begin{center}
\includegraphics[clip,bb= 3 21 564 527,width=0.65\linewidth]{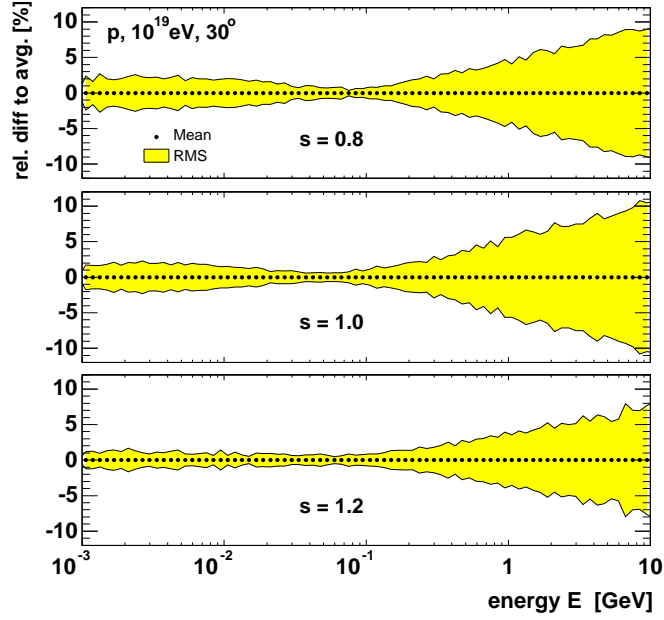}
\end{center}
\caption{Shower-to-shower fluctuations of electron energy spectra. The shaded bands indicate the RMS characterising 
the deviations of the individual showers of Fig.\,\ref{fig.ED_in_s} from the average.}
\label{fig.ED_in_s_reldiff2avg}
\end{figure}
In Figs. \ref{fig.ED_in_s} and \ref{fig.ED_in_s_reldiff2avg}, normalised differential electron 
energy spectra $f_{\rm e}~(X,E)$ of individual showers obtained with CORSIKA are compared for fixed (effective) 
shower age 
\begin{equation}
s=\frac{3}{(1+2X_{\rm max}/X)}~, 
\end{equation}
where $X$ denotes the slant depth (${\rm in~g/cm^{2}})$.
The study covers the shower age range $0.8 \le s \le 1.2$ that is most important for fluorescence observations. 
The spectra have been normalised according to
\begin{equation}
f_{\rm e}~(X,E)~=~\frac{1}{N_{\rm e}}\frac{{\rm d}N_{\rm e}}{{\rm d}\ln E}~,~~ 
{\rm with} ~~~~\int_{\ln E_{\rm cut}} f_{\rm e}~(X,E)~~{\rm d}\ln E = 1,
\label{eq.EDnorm}
\end{equation}
where $E_{\rm cut}$ is the energy threshold adopted in the simulation (1\,MeV in the examples shown).
The importance of the low-energy threshold $E_{\rm cut}$ for normalisation and Cherenkov (and fluorescence) light 
calculations is discussed in more detail below. 
If the longitudinal development is described in terms of shower age, the remaining shower-to-shower fluctuations are 
relatively small and due mainly to the depth of the first interactions. 
Deviations are smallest beyond the shower maximum because the individual distributions are the more
insensitive to initial differences in shower development the older the shower is. 
The small fluctuations of the envelopes in Fig.\,\ref{fig.ED_in_s_reldiff2avg} are of statistical kind 
and caused by the thinning applied in the simulations. 
The comparison is shown here only for proton showers of 
$10^{19}$\,eV for which the fluctuations are the largest and looks similar for different primary particle types at 
ultrahigh-energy ($>10^{17}$\,eV). 
%
%
%
%
\subsection{Universality}
\begin{figure}[t]
\begin{minipage}[c]{.48\linewidth}
\begin{center}
\includegraphics[clip,bb= 28 8 556 405,width=1.\linewidth]{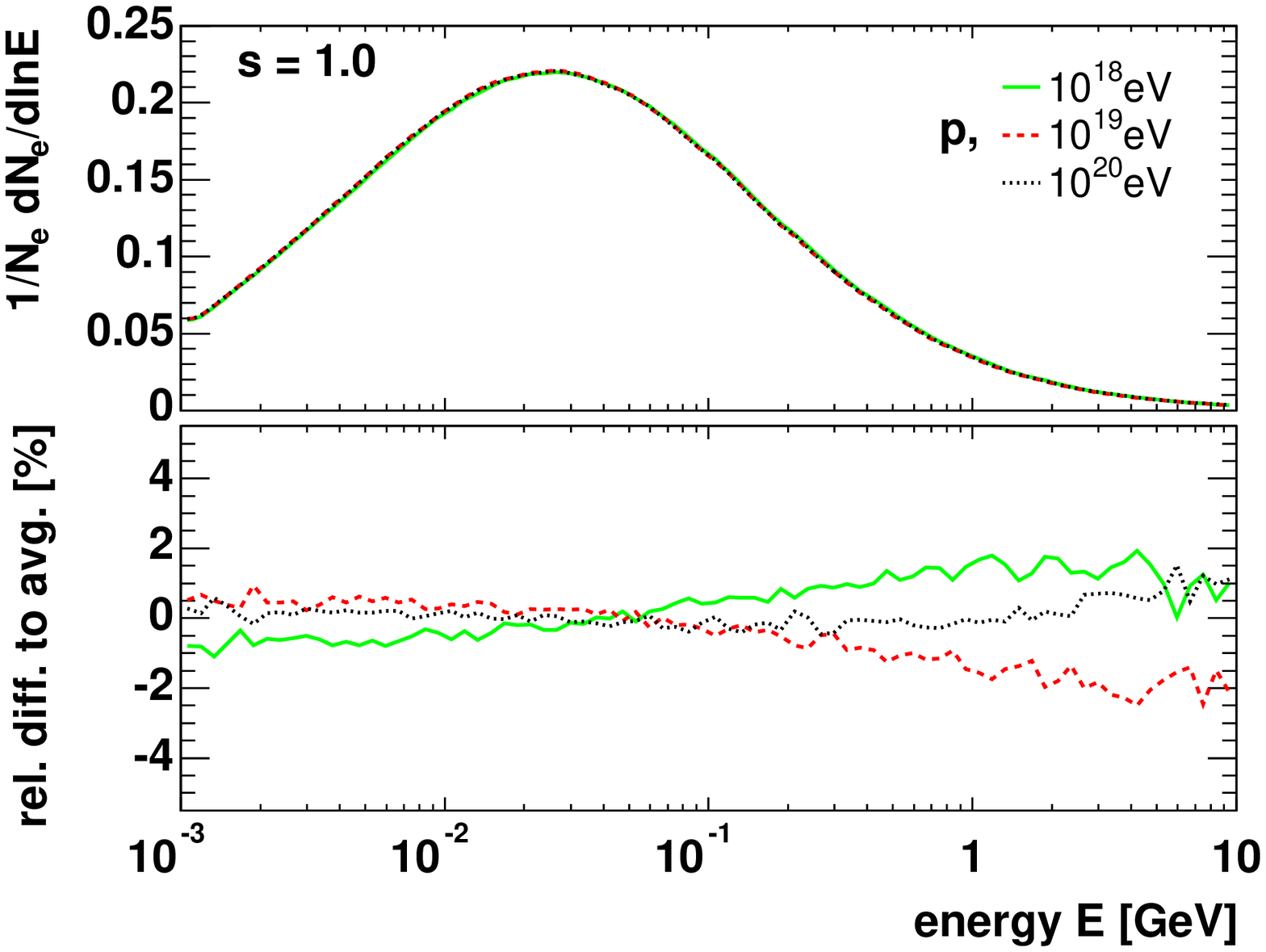}
\end{center}
\caption{Universality of electron energy spectra - different primary energy. 
In shower age, mean showers of $10^{18}, 10^{19}$ and $10^{20}$\,eV 
(each curve represents the mean distributions of 15 individual showers) do not show significant differences in their 
electron energy spectra; here shown for $s=1.0$ and proton initiated showers.}
\label{fig.ED_uni_E_p_s1}
\end{minipage}\hfill
\begin{minipage}[c]{.48\linewidth}
\begin{center}
\includegraphics[clip,bb= 28 8 556 405,width=1.\linewidth]{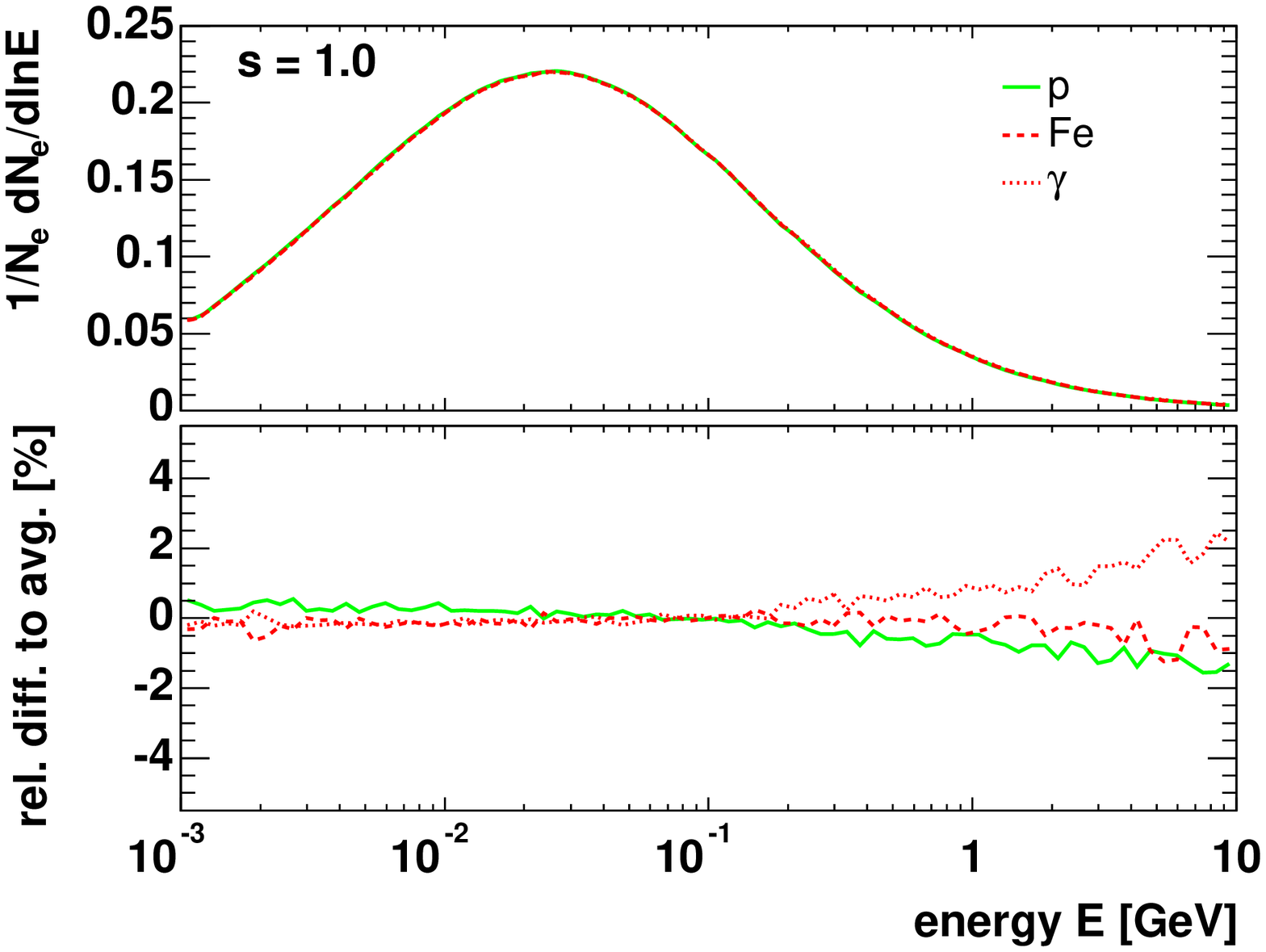}
\end{center}
\caption{Universality of electron energy spectra - different primary particle type. 
In shower age, mean proton, iron and gamma-ray 
showers of energies $> 10^{17}$\,eV (each curve represents the mean distribution 
averaged over $10^{18}, 10^{19}$, and $10^{20}$\,eV) 
do not show significant differences in their electron energy spectra, here shown for $s=1.0$.}
\label{fig.ED_uni_M_avg_s1}
\end{minipage}
\end{figure}
Average electron energy spectra $f_{\rm e}(E,s)$ for proton showers of different primary 
energies and showers of different primary particle types (proton, iron, and gamma-ray averaged over different 
energies $10^{18}$, $10^{19}$, and $10^{20}$\,eV) are shown\footnote{In the case of gamma-ray showers, 
simulations were done for $10^{19.5}$\,eV instead of $10^{20}$\,eV as otherwise 
primary gamma-rays interact with the Earth magnetic field well before reaching the atmosphere (pre-shower
effect) \cite{erber 1966}.} in Figs.\,\ref{fig.ED_uni_E_p_s1} and \ref{fig.ED_uni_M_avg_s1}. 
For the large electron energy range from 1\,MeV up to 10\,GeV, 
the mean distributions are similar when normalised according to (\ref{eq.EDnorm}). 
The energy distribution of electrons in UHECR initiated EAS 
does not depend significantly either on primary energy or on primary particle 
type, which allows a universal parameterisation in shower age \cite{giller 2004,nerling 2003}.
It is also independent of the shower zenith angle, however, 
this has only been investigated for inclinations smaller up to $60^{\circ}$. Increasing 
deviations from the spectral shape shown occur at $60^{\circ}$ for high-energy electrons in 
the GeV-range \cite{nerling 2005}. 
%
%
%
%
\subsection{Problem of low-energy cut and definition of shower size in simulations \label{SimDisk}}
\begin{figure}[t]
\begin{minipage}[c]{.48\linewidth}
\begin{center}
\includegraphics[width=0.85\textwidth]{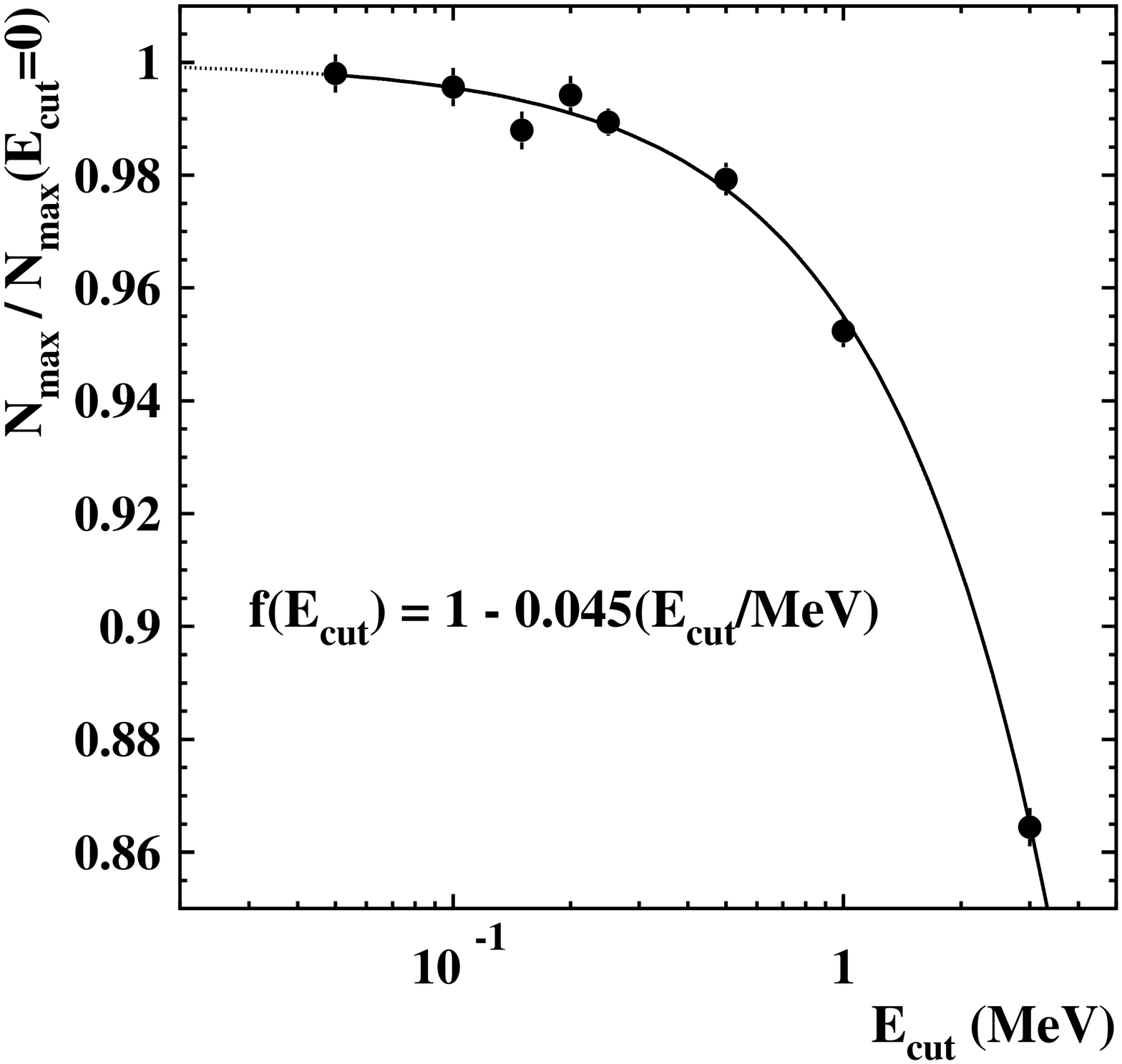}
\end{center}
\caption{Dependence of shower size at maximum on the simulation energy threshold. 
The ratio of the mean shower size at shower maximum $N_{\rm max}$ to that 
extra\-polated to $E_{\rm cut}=0$ is shown for different simulation thresholds. 
A parameterisation of this ratio as a function of $E_{\rm cut}$ is given \cite{nerling 2003}.}
\label{fig.Nmax_corr_Ecut}
\end{minipage}\hfill
\begin{minipage}[c]{.48\linewidth}
\begin{center}
\includegraphics[width=01.\textwidth]{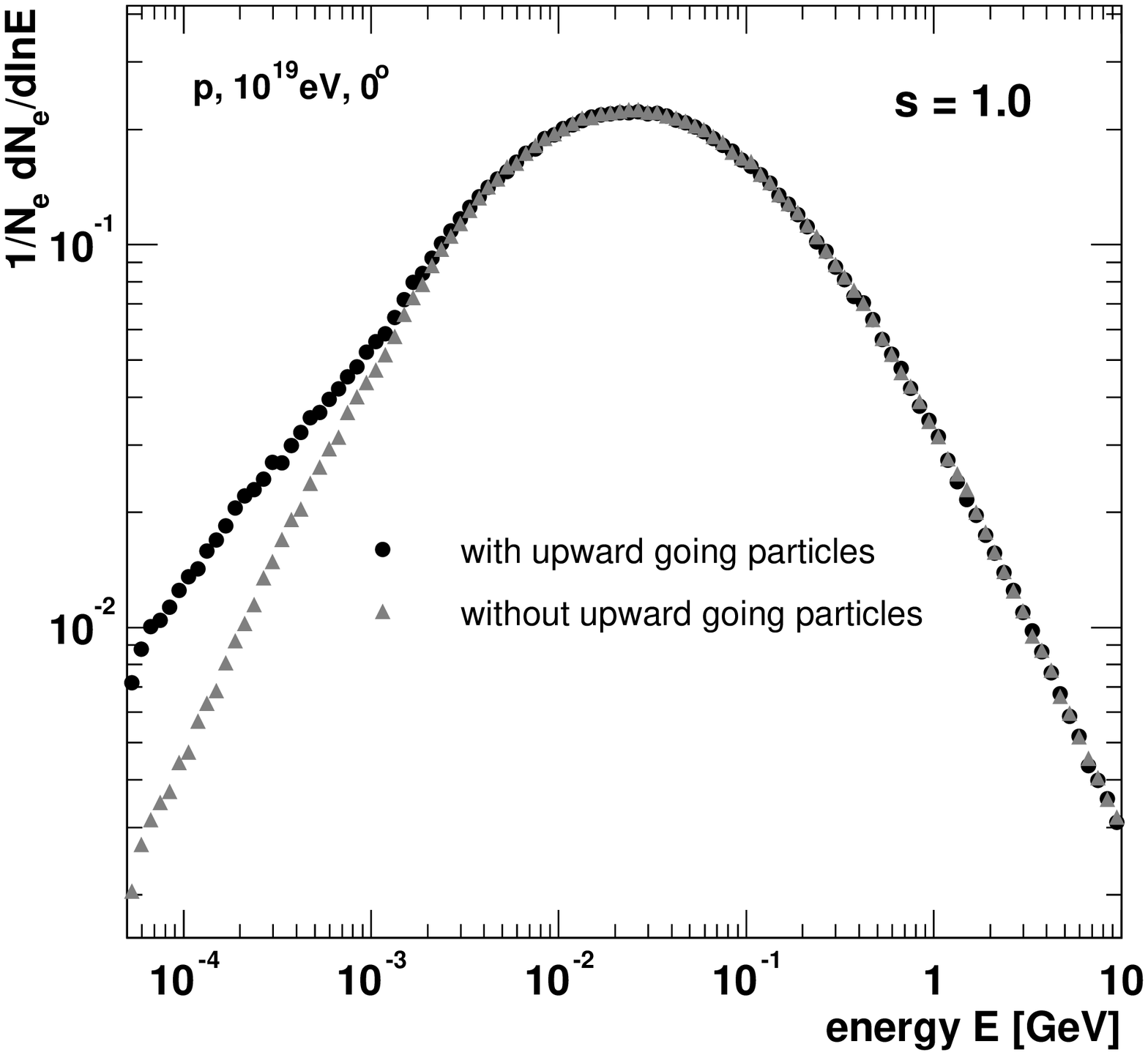}
\end{center}
\caption
{Electron energy spectra obtained from CORSIKA ($E_{\rm cut}=50$\,keV) with and without upward going particles. 
The spectra are normalised according to Eq.\,(\ref{eq.EDnorm}) assuming $E_{\rm cut}=1$\,MeV, for discussion see text.}
\label{fig.woupwup}
\end{minipage}
\end{figure}
In general, simulating air showers is not possible without applying a low-energy cut $E_{\rm cut}$ on the tracked
particles as the number of produced photons diverges for $E_{\rm cut}\rightarrow 0$ \cite{rossi 1941}.
The values of $E_{\rm cut}$ that are typically applied in simulations are in the range of 100\,keV - 3\,MeV. 
Therefore, a parameterisation of electron energy spectra for the purpose of 
calculating Cherenkov light from simulated shower size profiles has to account for different energy cuts to be 
consistent with the simulation.
Particularly in calculations of Cherenkov light based on ansatz (\ref{eq.CHprod}), 
and for deriving energy deposit profiles to compute fluorescence light production, 
the low-energy cut applied in simulations is important since the number of charged 
particles provided by the simulations refers only to the particles above the threshold $E_{\rm cut}$. 
The effect of different values of the simulation threshold on the shower size is illustrated in 
Fig.\,\ref{fig.Nmax_corr_Ecut}. 
For a detailed discussion of energy deposit calculations, see \cite{risse
2004,barbosa 2004}.

According to the definition of shower size, particles are counted when crossing virtual planes (horizontal in the CORSIKA version
used). Depending on the angular distribution, particles might not be counted when their tracks are 
parallel to these planes and when they are going upwards. Multiple scattering of particles might cause low-energy 
particles (back-scattered) being counted multiply, which implies some ambiguity in the definition of shower size. 
This ambiguity is mostly related to low-energy particles as can be seen in Fig.\,\ref{fig.woupwup}. The simulated 
electron energy spectrum of a vertical shower is shown for both upward going particles having been accounted for and 
not. Differences in the corresponding distributions occur for electrons below about 1\,MeV and increase with 
decreasing energy. 
Only electrons of less than about 1\,MeV are affected because of their angular distribution (with
respect to the shower axis), which is broad in the low-energy range up to a few MeV and steepens with 
increasing energy, see discussion in Sec.\,\ref{subsec.ADel}. 
%
%
%
%
\subsection{Parameterisation in shower age \label{ED}}
\begin{figure}[t]
\begin{center}
\includegraphics[clip,width=0.85\linewidth]{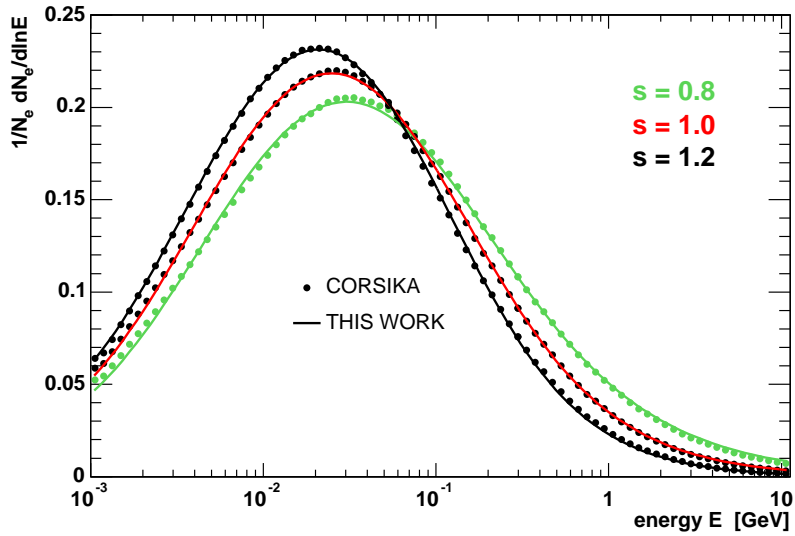}
\end{center}
\caption{Comparison of the new parameterisation Eq.\,(\ref{eq.EDpara}), see also \cite{nerling 2003}, 
and the energy spectra of an individual shower obtained with 
CORSIKA, proton, $10^{19}$\,eV.}
\label{fig.ED_MCvsNerl}
\end{figure}
Motivated by the high-energy limit of the energy behaviour of electrons in the cascade theory under 
approximation A \cite{rossi 1941} the following parameterisation 
\begin{equation}
f_{\rm e}~(E,s) ~=~a_{\rm 0}\cdot \frac{E}{(E+a_1)(E+a_2)^s}~, 
\label{eq.EDpara}
\end{equation}
is proposed.
As shown in Fig.\,\ref{fig.ED_MCvsNerl}, the CORSIKA spectra can be reproduced well
by ansatz (\ref{eq.EDpara}) using the parameters given in the appendix.
The parameter $a_0$ follows automatically from the normalisation condition (\ref{eq.EDnorm}).
The $E_{\rm cut}$-dependence of the normalisation $a_{0}$ is not negligible. In the energy range well 
below 250\,keV, this dependence might be negligible, but for larger cut-off values, the normalisation 
changes by up to e.g. about 10\,\% in the energy cut range from 50\,keV to 2\,MeV, 
see Fig.\,\ref{fig.Nmax_corr_Ecut}. A numerical expression for $a_0$ as a function of shower age
$s$ and $E_{\rm cut}$ is given in the appendix.

The parameterisation (\ref{eq.EDpara}) is compared with other parameterisations and electron energy spectra 
obtained with CORSIKA in the following. 
In the case of Hillas' parameterisation, the normalisation integral cannot be calculated directly.
Given the fact that this parameterisation was obtained for 100\,GeV primary photons, a
larger disagreement to CORSIKA above 15\,MeV (the lower validity limit given in
\cite{hillas 1982}) might have been expected.
Another functional form is proposed by Giller et al. in \cite{giller 2004}.
Neither Hillas' nor the parameterisation given by Giller et al. take into account different 
low-energy thresholds.
While in \cite{hillas 1982} $E_{\rm cut}=50$\,keV is given, no energy threshold has 
been specified in \cite{giller 2004}.
In the following, the parameterisations of Giller et al. is normalised the same 
way as the parameterisation introduced here, see Eq.\,(\ref{eq.EDnorm}), using $E_{\rm cut}=1$\,MeV. 
In the case of the integral distribution provided by Hillas, the parameterisation is renormalised 
to compare properly. In this case the renormalisation is done according to the CORSIKA results on the 
dependence of shower particle content on simulation energy threshold, as displayed in Fig.\,\ref{fig.Nmax_corr_Ecut}. 
\begin{figure}[t]
\begin{center}
\includegraphics[clip,width=0.85\linewidth]{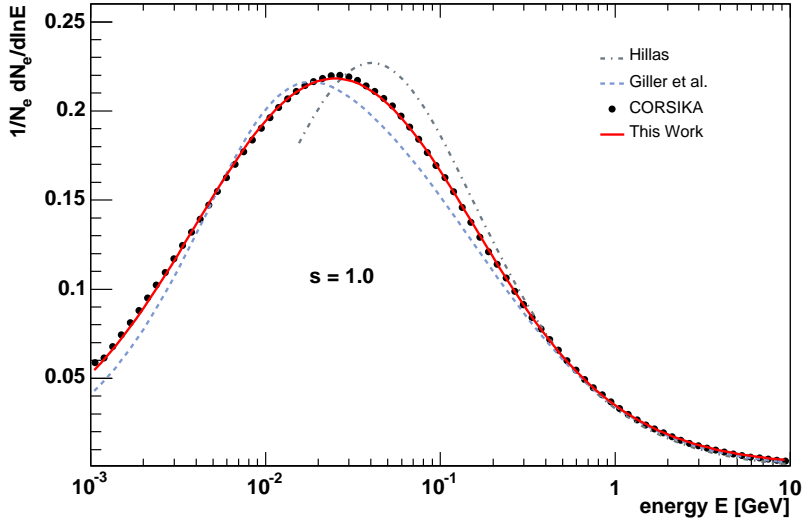}
\end{center}
\caption{Mean electron energy spectrum derived with CORSIKA (circles) for $s=1.0$. 
The Monte Carlo predictions are compared to the parameterisations according to Hillas 
\cite{hillas 1982} (dashed-dotted), the new parameterisation (Eq. \ref{eq.EDpara}) (solid) and Giller et al. 
\cite{giller 2004} (dashed).}
\label{fig.ED_MCvsAll}
\end{figure}
\begin{figure}[t]
\begin{center}
\includegraphics[clip,bb= 3 21 564 527,width=0.65\linewidth]{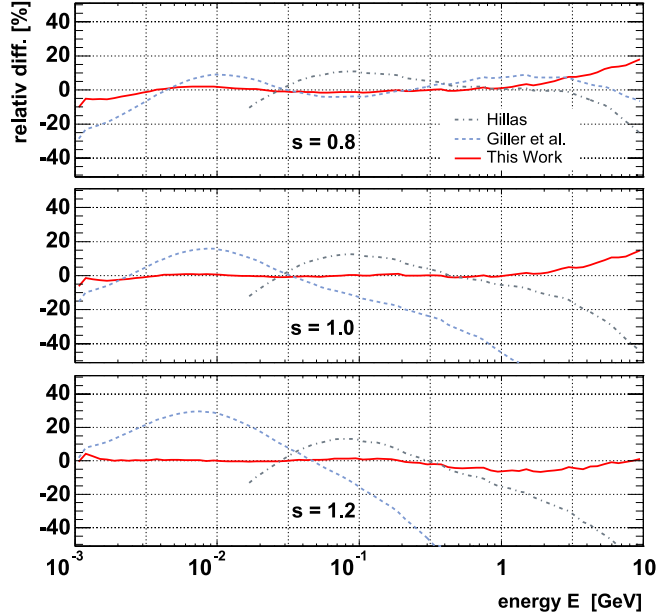}
\end{center}
\caption{Relative differences between model predictions and Monte Carlo results (relative to CORSIKA) 
as compared in Fig.\,\ref{fig.ED_MCvsAll} (\cite{hillas 1982} (dashed dotted), \cite{giller 2004} (dashed), and the 
parameterisation (Eq. \ref{eq.EDpara}) (solid)).}
\label{fig.ED_MCvsAll_reldiff}
\end{figure}

In Fig.\,\ref{fig.ED_MCvsAll} and \ref{fig.ED_MCvsAll_reldiff}, all three approaches
are compared to the CORSIKA predictions. The comparison is shown for a mean shower averaged in shower age 
over 87 individual showers of different primary energies and particle types ($10^{18}, 10^{19}$, and 
$10^{20}$\,eV; 45 proton, 12 iron, and 30 gamma-ray induced showers, cf. Fig.\,\ref{fig.ED_uni_E_p_s1} and 
Fig.\,\ref{fig.ED_uni_M_avg_s1}) at shower maximum. 
The other analytically calculated distributions show a remarkable shift of the maxima 
with respect to the CORSIKA result of about 25\,MeV at shower maximum (Hillas: $\sim$ 40\,MeV, 
Giller et al.: $\sim$ 20\,MeV), whereas the proposed parameterisation gives a very good 
overall description of the Monte Carlo results.
The relative differences of the different approaches with respect to CORSIKA are 
shown for three stages of shower development in Fig.\,\ref{fig.ED_MCvsAll_reldiff}.
The parameterisation (\ref{eq.EDpara}) predicts the CORSIKA results within a few percent for the whole
MeV-range, whereas uncertainties increase up to about 20\,\% at 10\,GeV. 
In particular the energy range most important for Cherenkov light calculations (20\,MeV to 1\,GeV) is 
described with high accuracy.
%
%
%
\subsection{Total number of Cherenkov photons produced \label{sec.CHtotnum}}
\begin{figure}[t]
\begin{center}
\includegraphics[clip,width=0.85\linewidth]{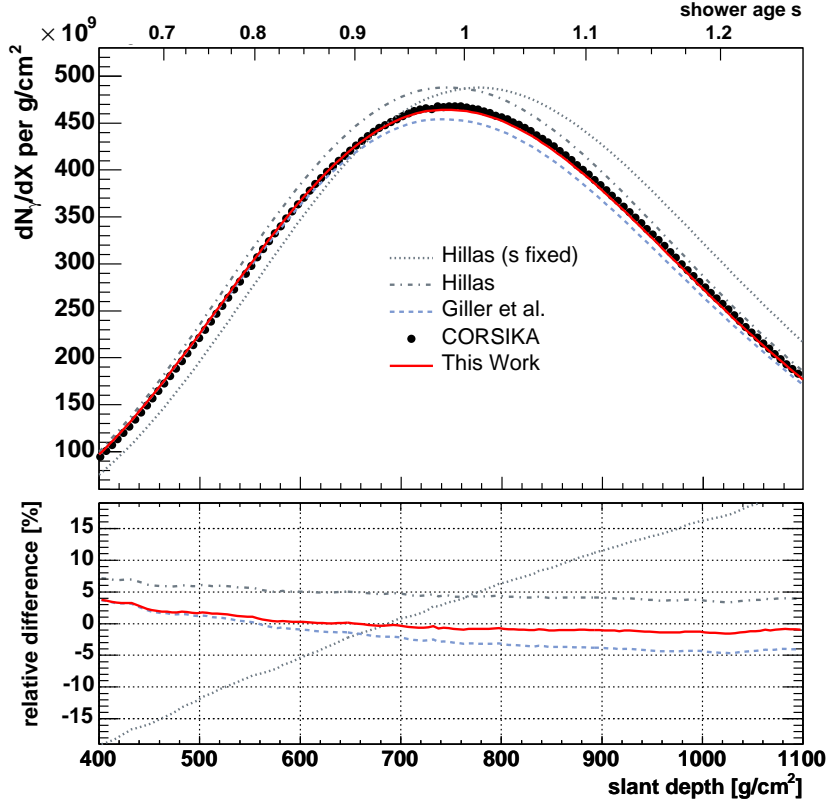}
\end{center}
\caption{Comparison of the total number of Cherenkov photons produced per slant depth within an
individual shower (proton, $10^{19}$\,eV, $30^{\circ}$) simulated with CORSIKA and analytically calculated by
ansatz (\ref{eq.CHprod}) using different parameterisations of electron energy
spectra. The different approaches are marked by dashed-dotted line for \cite{hillas 1982} 
and dotted line for the simplification $s=1$ as it has been used by the HiRes/Fly's Eye collaboration 
\cite{baltrusaitis 1985,abu-zayyad 2001}, dashed lines for \cite{giller 2004}, 
and solid lines for the one proposed in this paper. The accuracy of 
Eq.\,(\ref{eq.CHprod}) utilising parameterisation (\ref{eq.EDpara}) is better than 2\,\% over the whole 
range important for fluorescence observations.}
\label{fig.CHlong_all}
\end{figure}
According to ansatz (\ref{eq.CHprod}) the total number of produced Cherenkov photons can be
calculated applying the parameterisation (\ref{eq.EDpara}) of the electron energy
spectrum $f_{\rm e}(E,s)$ and the charged particle number $N_{\rm ch}(X)$ provided by CORSIKA. 
In Fig.\,\ref{fig.CHlong_all}, the result is
compared to the CORSIKA simulated profile and to analytical calculations based on the other
analytical formulae for electron spectra \cite{hillas 1982,giller 2004} already 
considered in the previous section.
Using the analytical expression going back to Hillas leads to an over-estimation of the 
Monte Carlo result, whereas the work of giller et al. \cite{giller 2004} results in an under-estimation. 
The calculation labelled `Hillas (s fixed)' employs the parameterisation of \cite{hillas 1982} for a fixed shower 
age of $s=1$ only, as often used in shower reconstruction (see e.g. \cite{baltrusaitis 1985}). This approximation leads to a shift of the
maximum of the Cherenkov profile by about 30-40 g/cm$^2$ towards larger depths, due mainly to the neglected
reduction of high-energy electrons with growing age.
Application of the parameterisation (\ref{eq.EDpara}) reproduces the CORSIKA simulated number 
of produced Cherenkov photons better than 1-2\,\% over the whole range important for fluorescence 
observations. 

The angular dispersion of charged particles, effectively increasing the Cherenkov yield per transversed depth d$X$ 
along the axis, has not been taken into account in the analytical approaches, cf. Sec.\,\ref{sec.CHprod}. 
Around the shower maximum, where the mean angle of electrons emitting Cherenkov light amounts about $8^{\circ}$, the 
corrected yield would increase the predicted curves by about +1\,\%. 
\clearpage

%
%
\section{Angular distribution of electrons and Cherenkov photons \label{sec.AD}}
%
%
%
%
\subsection{Electron angular distribution \label{subsec.ADel}}
The energy dependent electron angular distribution with respect to the shower axis implies the angular distribution 
of resultant Cherenkov photons produced in EAS by convoluting in the Cherenkov emission angle. 
This angle slightly changes with altitude and amounts about $1.4^{\circ}$ at 
sea level and decreases to about $0.8^{\circ}$ at 10\,km above sea level.
Electrons undergo Coulomb scattering and the higher the particle energy the smaller is the 
scattering angle. Thus, the electron angular distribution is correlated with energy: 
The higher the mean electron energy, the smaller is the mean angle with respect to the shower axis.

\begin{figure}[t]
\begin{center}
\includegraphics[clip, width=1.\linewidth]{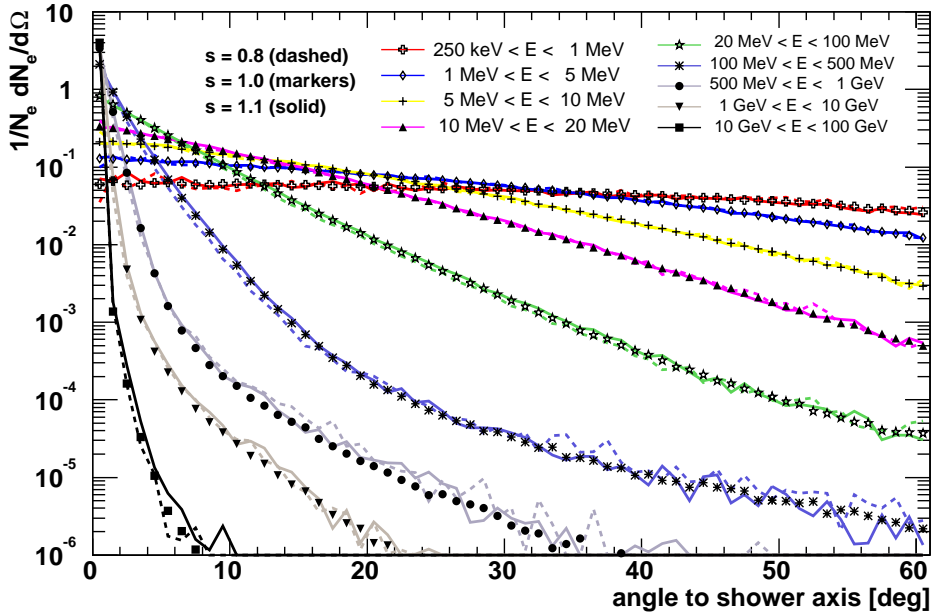}
\end{center}
\caption{Normalised electron angular distributions for an individual vertical proton shower of 
$10^{19}$\,eV. 
The electron angular distribution (with respect to shower axis) is shown 
for three different shower ages and various different ranges of electron energies. 
The electron angular distribution in high-energy showers is to a large extend independent of the shower age.}
\label{fig.ADel_age}
\end{figure}
In Fig.\,\ref{fig.ADel_age}, the normalised electron angular 
distribution with respect to the shower axis (averaged over azimuthal angle) is shown exemplarily for an individual 
proton shower of $10^{19}$\,eV for three different shower ages, and various 
ranges of electron energies. 
The distributions do not change significantly with shower age within the statistical limitations. 
Deviations increase with increasing angles 
to the shower axis, especially for higher electron energies of a few hundred MeV and GeV. 
However, at that energies and angles, fluctuations are also caused by low statistics 
due to the thinning applied. 
That the electron angular distributions is practically independent of shower age supports the electrons energy to 
be the determining factor of their angular distribution.

Charged particles are also deflected by the geomagnetic field. 
In the following we always average over the azimuthal angle and do not consider asymmetries implied by the 
geomagnetic field. Although such effect has been discussed not to make a gross change \cite{hillas 1982}, the 
effect seems not to be negligible, an approximate treatment of the effect is given in \cite{elbert 1983}. 
At which extend the formulae given there describes the resulting azimuthal asymmetry sufficiently well 
would have to be investigated in detail separately.
%
%
%
%
\begin{figure}[t]
\begin{center}
\includegraphics[clip,width=1.\linewidth]{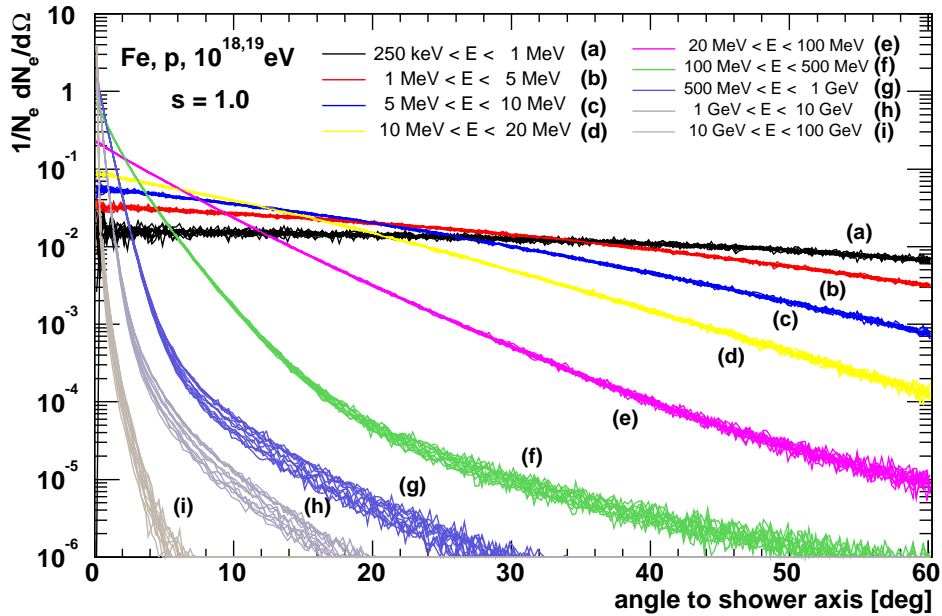}
\end{center}
\caption{Universality of (normalised) electron angular distributions. The electron angular distribution with 
respect to shower axis of numerous individual showers of different energies and primary particles is shown 
at shower maximum for various different ranges of electron energies. 
(proton, iron, $10^{18}$\,eV, $10^{19}$\,eV, $0^{\circ}$).}
\label{fig.ADel_uni}
\end{figure}
\subsection{Universality}
As electron energy spectra in high-energy showers are universal and their scattering 
angle is mostly determined by the particle energy, the electron angular distribution should 
also be universal.

In Fig.\,\ref{fig.ADel_uni}, the electron angular distribution of many individual proton and 
iron showers is shown for a fixed shower age ($s=1.0$). For the MeV-range, the distributions of 
individual showers (of different primary energies and primary particle types) do not differ 
much; larger fluctuations occur only for large angles. The angular distribution 
of electrons in the GeV-region shows a larger spread. This is in agreement with the universality 
studies of the energy spectrum, where larger shower-to-shower fluctuations also occur in the 
GeV-range.

In conclusion, the spectral shape of the angular distribution of electrons in high-energy showers 
is mostly determined by the electron energy and almost independent of the shower age (in the range
considered of $0.8<s<1.2$). It does not depend significantly either on primary energy 
or on primary particle type, see also \cite{nerling 2005_a,giller 2004b}. 
%
%
%
%
\begin{figure}[t]
\begin{center}
\includegraphics[clip,width=1.0\linewidth]{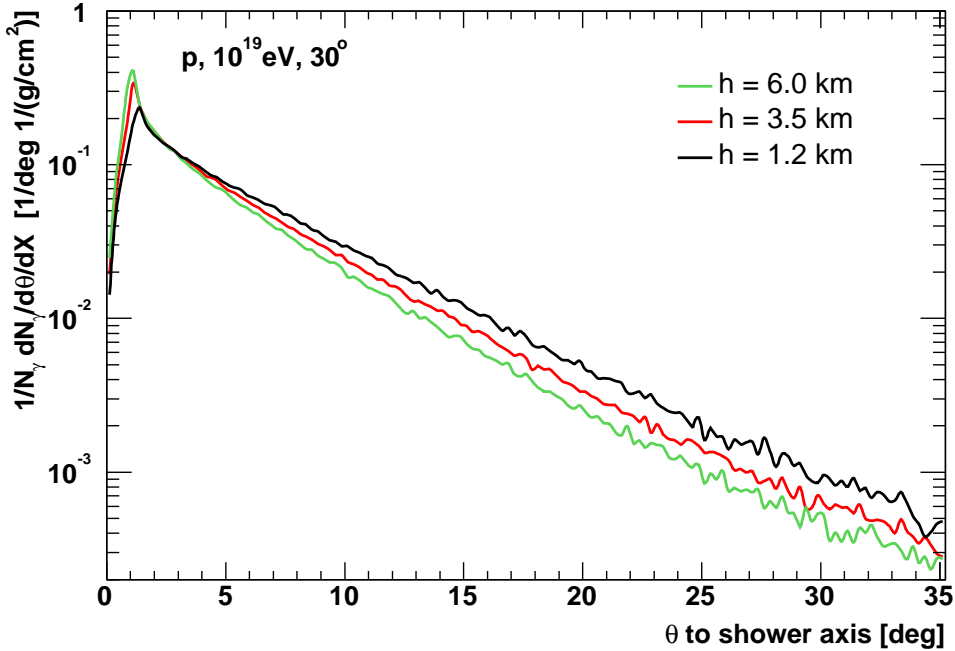}
\end{center}
\caption{The angular distribution of Cherenkov photons with respect to the shower axis 
for 3 different heights obtained with CORSIKA is shown for one individual proton shower.}
\label{fig.AD3heights}
\end{figure}
\subsection{Resultant Cherenkov photons}
The angular distribution $A_{\gamma}(\theta,X,h)$ in Eq.\,(\ref{eq.CHprod}) is the angular distribution of Cherenkov photons produced
per angular bin with respect to the shower axis normalised to one photon (averaged over azimuth):
\begin{equation}
A_{\gamma}(\theta,X,h)\cdot\frac{{\rm d}N_{\rm \gamma}}{{\rm d}X}~=~\frac{{\rm d}N_{\rm \gamma}}{{\rm d}\theta{\rm d}X}~(\theta,X,h)~,
{\rm ~with} \int_0^\pi A_{\gamma} ~{\rm d}\theta ~=~ 1~.
\label{eq.ADdef}
\end{equation}
$A_{\gamma}(\theta)$ follows from the angular distribution of underlying electrons.
Thus, two dependencies follow for the angular distribution of produced Cherenkov photons in EAS:
\begin{itemize}
\item[(i)] \textit{Dependence on height} 
\newline
The Cherenkov emission angle is slightly changing with altitude, this dependence is, however, negligible 
as the electron angles are much larger. 
Only electrons of energies above the Cherenkov energy threshold $E_{\rm thr}$ 
are of interest, which implies that the higher $E_{\rm thr}$ the
smaller is the mean angle of electrons emitting Cherenkov light.
Finally, since the Cherenkov energy threshold $E_{\rm thr}(h)$ is - via the refractive index $n(h)$
- a function of altitude, also the angular distribution of Cherenkov
photons depends on height. This dependence on $E_{\rm thr}$ can be approximated by an exponential function, see
\cite{elbert 1983}.
\\
\item[(ii)] \textit{Dependence on shower age} 
\newline
Since electron energy spectra develop with shower age, the part of electrons above the Cherenkov 
threshold energy $E_{\rm thr}$ also changes with shower age. This is the reason why also the angular distribution 
of Cherenkov photons depends on shower age.
\end{itemize}
\begin{figure}[t]
\begin{center}
\includegraphics[clip,width=1.\linewidth]{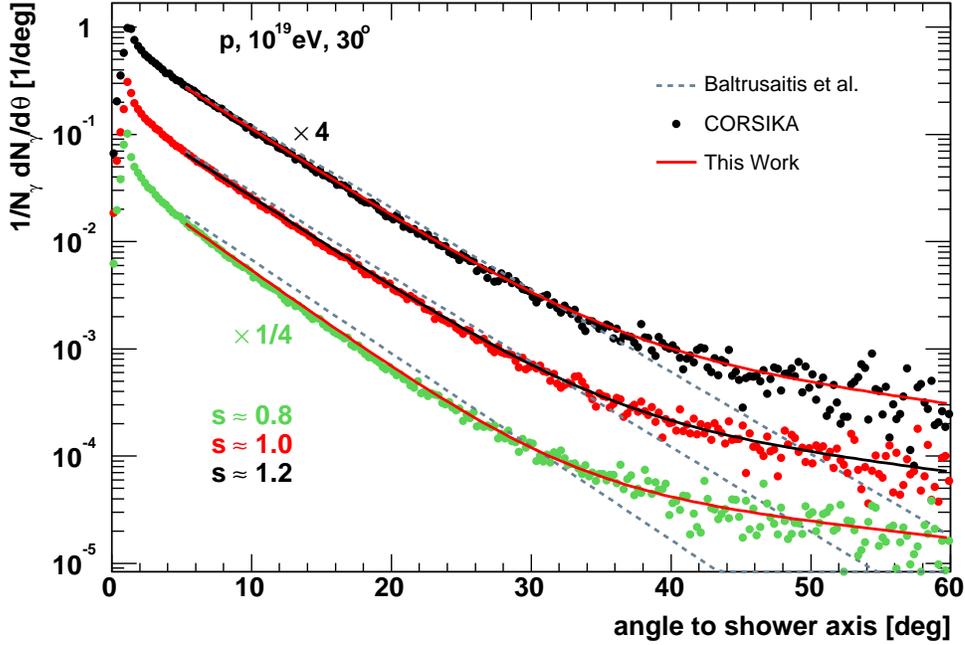}
\end{center}
\caption{Angular distribution of produced Cherenkov photons with respect to the shower axis 
in a single CORSIKA shower for $s=0.8,1.0,$ and 1.2. The Monte Carlo results are compared to 
predictions from \cite{baltrusaitis 1985} and the new
parameterisation, Eq.\,(\ref{eq.ADpara}).}
\label{fig.ADcheck_p19th30}
\end{figure}
Consequently, the photon angular distributions depend on both, the height 
due to the dependence of $E_{\rm thr}(h)$, and on the shower age because of 
the dependence of the electron energy spectrum $f_{\rm e}(E,s)$. 
These dependencies of the angular distribution of produced Cherenkov photons can be seen in 
Fig.\,\ref{fig.AD3heights}. With decreasing height and increasing index 
of refraction, the photon number at larger angles to the shower axis increases, whereas it decreases at 
angles smaller than about $2.5^{\circ}$, see also Fig.\,\ref{fig.ADcheck_p19th30}.
%
%
%
%
\subsection{Parameterisation in height and shower age}
In principle, one could use the universal electron angular distribution (finding a parameterisation in the 
mean electron energy or just use tabulated mean values) for calculating the
angular distribution of produced Cherenkov photons. As it turns out it is also possible to find a closed
analytical approximation for the complete integral.

It is common to describe the height dependence of the angular distribution of produced Cherenkov photons
by an exponential function, where the scaling angle $\theta_{\rm 0}$ is a function of the Cherenkov energy 
threshold, see e.g. \cite{baltrusaitis 1985,elbert 1983}:
\begin{equation}
A_{\gamma}(h,\theta) = 1/\theta_{\rm 0}\cdot e^{-\theta/\theta_{\rm 0}}~.
\label{eq.A_h}
\end{equation}
Parameterisations of $\theta_{\rm 0}$ as a function of local Cherenkov threshold energy $E_{\rm thr}(h)$
have been calculated by several authors; for example $\theta_{0}=a\,E_{\rm thr}^{-b}\,$, 
with $(a,b)=(0.83, 0.67)$ \cite{stanev 1981}.
\begin{figure}[t]
\begin{center}
\includegraphics[clip,bb= 31 48 561 597,width=0.85\linewidth]{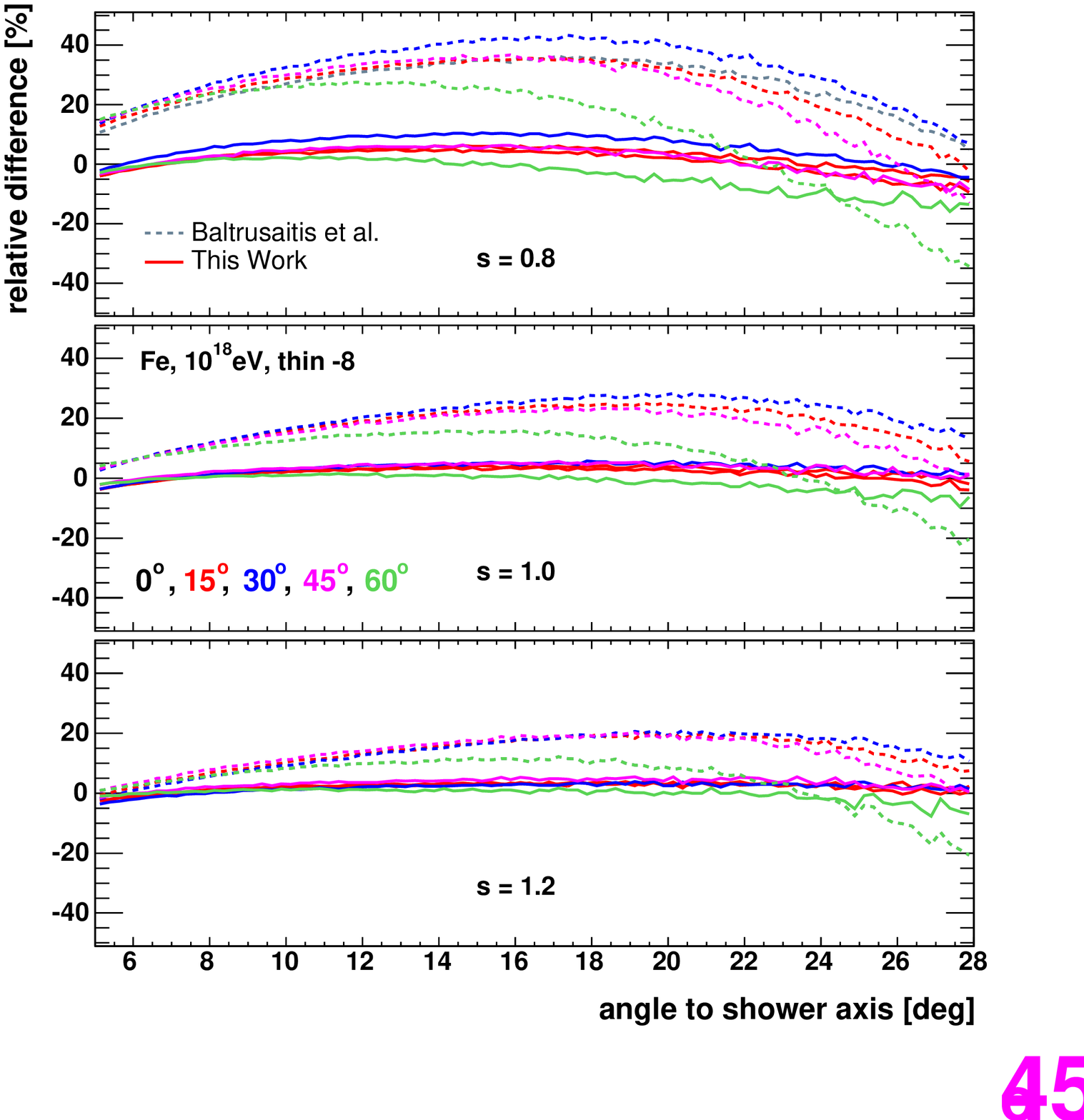}
\end{center}
\caption{Quality of description of individual CORSIKA showers (iron of $10^{18}$, different
zenith angles) by the new parameterisation
Eq.\,(\ref{eq.ADpara}). The statistical fluctuations due to thinning are reduced remarkably as 
very good thinning is applied ($10^{-8}$). Showers are described practically independently of their inclinations.}
\label{fig.AD_sh2para_thin-8}
\end{figure}
Traditionally this approximation is applied for calculating the Cherenkov contamination of fluorescence 
light signals from high-energy showers, see e.g. \cite{baltrusaitis 1985,abu-zayyad 2001}.
Generalising this ansatz to take into account both effects, the dependence on 
refractive index as well as on the shower age we, write
\begin{eqnarray}
   A_{\gamma}(\theta,h,s)~~&=&~~A_{\gamma}(s)\cdot A_{\gamma}(\theta,h)~~ \nonumber\\
   &=&~~a_{\rm s}(s)~\cdot~1/\theta_{\rm c}(h)~e^{-\theta/\theta_{\rm c}(h)}~, 
\label{eq.ADansatz}
\end{eqnarray}
where $a_{\rm s}(s)$ is a polynomial of second order in shower age and 
the second, exponential term on the right depends on altitude only. 
The first accounts for so-called shower-to-shower fluctuations, namely the position of $X_{\rm max}$, 
and the latter includes the inhomogeneity of the medium, namely the refractive index changing with altitude.
The most important range up to about $30^{\circ}$ is described well by Eq.\,(\ref{eq.ADansatz}). 
To enlarge the range of validity up to $60^{\circ}$ and improve the data description around $30^{\circ}$, 
ansatz (\ref{eq.ADansatz}) is extended by a second term to
\begin{eqnarray}
A_{\gamma}(\theta,h,s)~~=~~a_{\rm s}(s)~\frac{1}{\theta_{\rm c}(h)}~e^{-\theta/\theta_{\rm c}(h)}~
+~b_{\rm s}(s)~\frac{1}{\theta_{\rm cc}(h)}~e^{-\theta/\theta_{\rm cc}(h)}~.
\label{eq.ADpara}
\end{eqnarray}
The numerical values of $a_{\rm s}$, $b_{\rm s}$, and $\theta_{\rm c}$, $\theta_{\rm cc}$, which have been found by a 
global fit to many individual showers of different primary particles and inclinations, are given in the
appendix. The simulated Cherenkov photon angular distributions, which have been used to determine the coefficients
of (\ref{eq.ADpara}), although simulated with geomagnetic field, are averaged over
azimuth. 
As shown in Fig.\,\ref{fig.ADcheck_p19th30}, the CORSIKA spectra are described properly by ansatz 
(\ref{eq.ADpara}). 
The achieved quality of description by this ansatz is shown in Fig.\,\ref{fig.AD_sh2para_thin-8}. 
There, the relative differences with respect to individual CORSIKA showers of very high statistics (optimum 
thinning $10^{-8}$) are shown. 
Taking into account both the height and age dependence, also showers 
of different inclinations are described practically with the same accuracy of a few percent.

%
%
\section{Use of energy deposit for reconstruction and simulation of light profiles \label{sec.Edep}}

In fluorescence observations, one measures the fluorescence light profile, and the number of
produced fluorescence photons in a shower is proportional to the local ionisation energy deposit 
d$E_{\rm dep}/{\rm d}X$:
\begin{equation}
\label{eq.Flprod}
\frac{{\rm d}N_{\gamma}^{\rm fl}}{{\rm d}X}(X)
~=~y_{\gamma}^{\rm fl}(h) \cdot \frac{{\rm d}E_{\rm dep}}{{\rm d}X}(X)~,
\end{equation}
where $y_{\gamma}^{\rm fl}$ is the fluorescence yield in air at altitude $h$. 
Thus, light profiles measured using the fluorescence technique are measurements of the
energy deposit by the shower rather than its shower size.
The advantage of using d$E_{\rm dep}/{\rm d}X$ over shower size is not only that 
d$E_{\rm dep}/{\rm d}X$ is most closely connected to the measured fluorescence light but also solves the 
problem of the ambiguity in the definition of shower size in simulations as pointed out in Sec.\,\ref{sec.ED}. 
Furthermore, it can be conveniently simulated \cite{risse
2004,barbosa 2004}. In addition, energy deposit profiles can be simulated with rather large low-energy 
thresholds $E_{\rm cut}\ge 1$\,MeV, and therefore is much less CPU consuming.

For analytical description of the Cherenkov contribution to fluorescence profiles, it is hence very useful 
to allow the calculation additionally as a function of d$E_{\rm dep}/{\rm d}X$, although 
the shower size is the physics quantity that is more closely related to Cherenkov production.
%
%
%
%
\subsection{Parameterisation of mean ionisation loss rate}
\begin{figure}[t]
\begin{center}
\includegraphics[clip,bb= 49 49 503 373, width=0.85\linewidth]{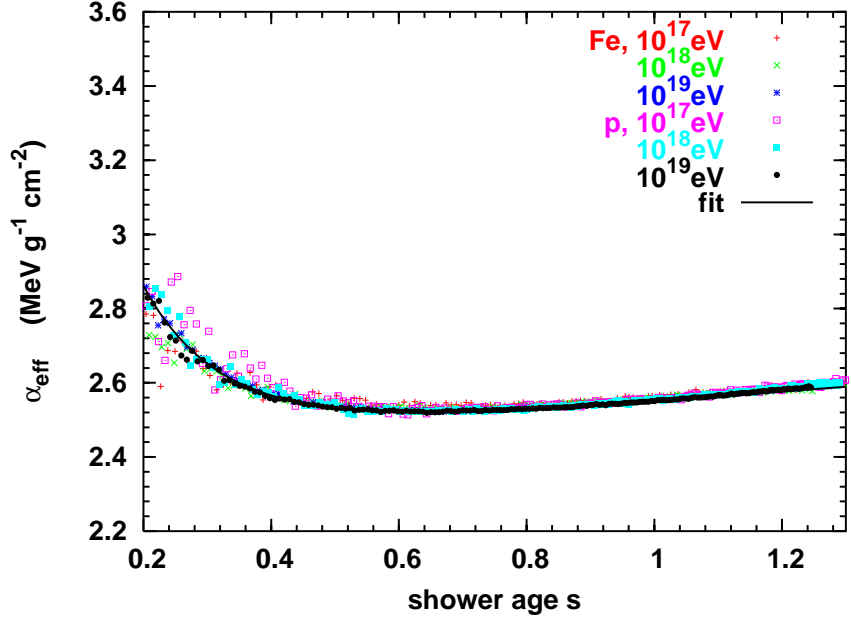}
\end{center}
\caption{The Mean ionisation loss rate $\alpha_{\rm eff}$ as obtained from CORSIKA simulations according to 
Eq.\,(\ref{eq.alpha_eff}) is shown for vertical showers of different primary energies and particles 
($10^{17}, 10^{18}$, $10^{19}$\,eV, and proton, iron). 
The fit shown is given by Eq.\,(\ref{eq.alpha_para}). 
}
\label{fig.alpha_eff}
\end{figure}
We define the mean ionisation loss rate $\alpha_{\rm eff}$ by
\begin{equation}
\label{eq.alpha_eff}
\alpha_{\rm eff}~(X, E>E_{\rm cut}) ~ N_{\rm ch}~(X, E>E_{\rm cut}) ~=~\frac{{\rm d}E_{\rm dep}}{{\rm d}X}~(X)~,
\end{equation} 
where $E_{\rm cut}$ is a low-energy threshold, which has to be applied in the case of 
shower simulations as discussed in Sec.\ref{sec.ED}. 
The ionisation loss rate is mostly determined by the charged particle energy. 
As a consequence of the universality of electron distributions, cf. Sec.\ref{sec.ED} and \ref{sec.AD}, 
also $\alpha_{\rm eff}$ should neither depend on primary energy nor on particle type. 

The corresponding study is shown in Fig.\,\ref{fig.alpha_eff} where 
$\alpha_{\rm eff}$ obtained from CORSIKA simulations is shown for individual vertical proton and iron showers of 
different energies ($10^{17}$, $10^{18}$, $10^{19}$). 
In the simulations shown, $E_{\rm cut}=1$\,MeV has been applied. 
For the large range from about $s=0.5$ to 1.2, the mean ionisation loss rate $\alpha_{\rm eff}(s)$ does not differ 
significantly for different primary particles and energies. The shower age dependence of $\alpha_{\rm eff}(s)$ can be described by 
\begin{equation}
\label{eq.alpha_para}
\alpha_{\rm eff}~(s) ~=~\frac{c_1}{(c_2 + s)^{c_3}} + c_4 + c_5 \cdot s~,
\end{equation}
which is the analytical expression superimposed in Fig\,\ref{fig.alpha_eff}; 
the parameter values are given in the appendix.
%
%
%
%
\subsection{Calculation of Cherenkov light as a function of energy deposit}
\begin{figure}[t]
\begin{center}
\includegraphics[clip,width=0.8\linewidth]{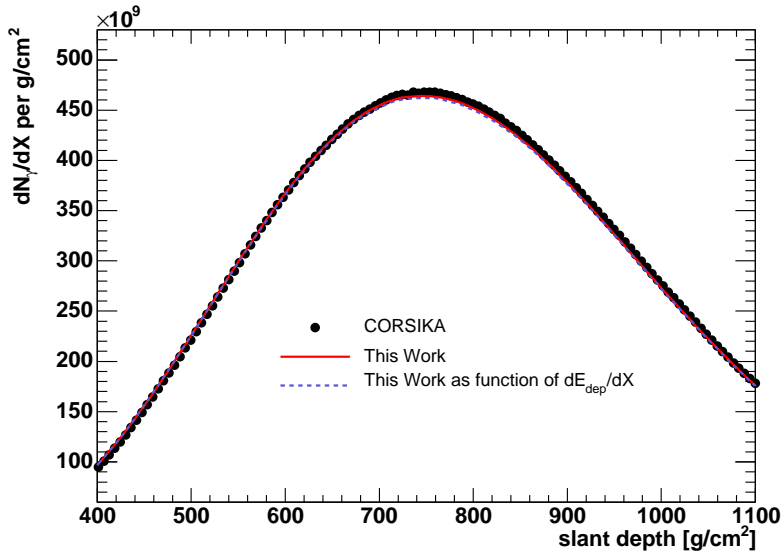}
\end{center}
\caption{Comparison of total number of Cherenkov photons produced per slant depth within an
individual shower (proton, $10^{19}$\,eV, $30^{\circ}$). 
Shown are the results of the CORSIKA simulation and analytically calculated by
ansatz (\ref{eq.CHprod}) and (\ref{eq.CHprod_Edep}) respectively using the proposed parameterisation of electron energy
distributions (Eq.\,(\ref{eq.EDpara})).}
\label{fig.Ch_longEdep}
\end{figure}
Using parameterisation (\ref{eq.alpha_para}) in shower age $s$, Cherenkov light production can be
calculated as a function of energy deposit profiles following ansatz (\ref{eq.CHprod})
\begin{eqnarray}
\label{eq.CHprod_Edep}
\frac{{\rm d}N_{\gamma}}{{\rm d}X}(X,h)~&=&~N(X)
\int_{\ln E_{{\rm thr}}}y_{\gamma}(h,E)~f_{\rm e}~(s,E)~{\rm d}\ln E \nonumber\\
&=&~\frac{1}{\alpha_{\rm eff}(s)}~\frac{{\rm d}E_{\rm dep}}{{\rm d}X}(X)
\int_{\ln E_{{\rm thr}}}y_{\gamma}(h,E)~f_{\rm e}~(s,E)~{\rm d}\ln E~.
\end{eqnarray}
However, it should be noted that for proper calculations, $f_{\rm e}(s,E>E_{\rm cut})$ and 
$\alpha_{\rm eff}(s,E>E_{\rm cut})$ (and $N(X, E>E_{\rm cut}$) have to refer to the same low-energy 
threshold applied in the simulation. 
Moreover, vice versa, one can derive the energy deposit profile from a given shower size profile 
applying the parameterisation (\ref{eq.alpha_para}). 
Such application is useful e.g. for simulating fluorescence profiles by so-called
fast hybrid simulation codes like e.g. CONEX \cite{kalmykov 2003}, which give shower size profiles as output.

The Cherenkov calculation according to Eq.\,(\ref{eq.CHprod_Edep}) is shown in Fig.\,\ref{fig.Ch_longEdep}
and compared to the CORSIKA result as well as to the calculation as a function of shower size. 
The same accuracy in reproducing the full Monte Carlo calculation better than 2\,\% is achieved either using the shower size 
or the energy deposit profile provided by the simulations.

%
%
\section{Summary and conclusions}
\label{Sum}
For the purpose of developing an analytical description of Cherenkov light production in EAS, 
the universality of electron distributions in high-energy showers has been investigated, 
namely the energy and angular distribution. 
In shower age, both have been shown to be independent of different primary energy and particle 
type to a good approximation. 
A parameterisation of the energy spectrum in shower age has been
introduced, which describes showers independently of different primary energy, particle type and zenith angle 
at high accuracy of a few percent (within shower-to-shower fluctuations) for the electron energy range from 1\,MeV to a few GeV, 
covering also the range most important for Cherenkov light emission. 

The electron angular distribution is mostly dominated by the particle energy and do not depend
significantly on the shower age for a fixed energy. As a consequence, the angular distribution of
produced Cherenkov photons has been parameterised as a function of height and shower age. 
The age dependence follows from the electron energy spectra. The dependence on height is due to 
the change of the Cherenkov threshold energy with altitude. 

Based on these universality features of high-energy showers, an analytical description of the Cherenkov light 
production in EAS has been presented providing both, the total number of produced Cherenkov photons as well as their angular 
distribution with respect to the shower axis. It offers the calculation of the direct and scattered
Cherenkov contribution to measured fluorescence light profiles, see \cite{nerling 2005_b}.

The advantage of using energy deposit rather than the number of charged particles in the simulation and reconstruction of 
light profiles measured with the fluorescence technique has been pointed out. 
Having introduced a parameterisation of the mean energy deposit $\alpha_{\rm eff}\,(s)$, 
the scattered and direct Cherenkov light contributions to light profiles measured in fluorescence observations 
can be estimated either as a function of shower size or energy deposit profiles 
with an accuracy of a few percent. 

The geomagnetic effect leading to azimuthal asymmetries in electron and Cherenkov photon
angular distributions seems not to be negligible and needs to be studied in more detail separately.

%
%
\section{Acknowledgements}
We thank our colleagues of the Auger Collaboration for many fruitful discussions. 
Particularly, the support of D. Heck in performing the simulations and sharing his deep insight in 
CORSIKA is gratefully acknowledged. 
F. Nerling acknowledges the receipt of travel support by the Deutsche Forschungsgemeinschaft via the
`Graduiertenkolleg Hochenergiephysik und Teilchenastrophysik' of the University of Karlsruhe.
\label{Ack}


\newpage

%
%
\section{Appendix}
%
%
\subsection{Parameterisation of electron energy spectrum}
The parametrisation of the normalised electron energy spectrum in EAS is given by 
\begin{equation}
\label{eq.EDpara_app}
f_{\rm e}~(E,s) ~=~ a_0 ~\cdot~ \frac{E}{(E+a_1)~(E+a_2)^s}~.
\end{equation}
The CORSIKA simulated electron energy spectra are described well by the
following set of parameters:
\begin{eqnarray}
a_1 &=& 6.42522 - 1.53183 \cdot s  \nonumber\\
a_2 &=& 168.168 - 42.1368 \cdot s~, {\rm with~} E~{\rm in}~{\rm MeV}.     
\end{eqnarray}
For applications in data analysis, the normalisation $a_0$ 
has been parametrised in shower age $s$ for different $E_{\rm cut}$. 
The parameter $a_0$ is described by an exponential function in shower age
\begin{equation}\label{eq.para_a0}
 a_0 = k_0 \cdot \exp( k_1 \cdot s + k_2 \cdot  s^2)~,
\end{equation}
where the parameters $k_0, k_1$ and $k_2$ calculated for six different threshold energies in the typical 
range for simulation energy thresholds of 50\,keV~- 2\,MeV
may be linearly interpolated from the tabulated values given in 
Tab.\,\ref{tab.para_a0} .
\begin{table}[b]
 \caption{Tabulated values for the normalisation $a_0$ (see Eq.\,(\ref{eq.para_a0})) of electron energy 
 parameterisation (\ref{eq.EDpara_app})
\label{tab.para_a0}
 }
\begin{center}
\begin{tabular}{|c||c|c|c|} 
\hline  
$E_{\rm cut}$ [MeV]  & $k_0$  & $k_1$  & $k_2$        \\ \hline
2.	& 1.48071e-01& 6.22334& -5.89710e-01 \\ \hline
1.	& 1.45098e-01& 6.20114& -5.96851e-01 \\ \hline
0.5	& 1.43458e-01& 6.18979& -6.01298e-01 \\ \hline
0.25    & 1.42589e-01& 6.18413& -6.03838e-01 \\ \hline
0.1     & 1.42049e-01& 6.18075& -6.05484e-01 \\ \hline
0.05    & 1.41866e-01& 6.17963& -6.06055e-01 \\ \hline
 \end{tabular}
\end{center}
\vspace{-0.5cm}
\end{table}

%
%
\subsection{Parameterisation of angular distribution of Cherenkov photons}
The parameterisation of the angular distribution of produced Cherenkov photons with respect to the shower axis 
is given by 
\begin{eqnarray}
A_{\gamma}(\theta,h,s)~~=~~a_{\rm s}(s)~\frac{1}{\theta_{\rm c}(h)}~e^{-\theta/\theta_{\rm c}(h)}~
+~b_{\rm s}(s)~\frac{1}{\theta_{\rm cc}(h)}~e^{-\theta/\theta_{\rm cc}(h)}
\label{eq.ADpara_app}
\end{eqnarray}
In this expression the age dependence is included by
\begin{eqnarray}
a_{\rm s}(s) &=& a_{\rm 0} + a_{\rm 1} \cdot s + a_{\rm 2}\cdot s^{2} \\
b_{\rm s}(s) &=& b_{\rm 0} + b_{\rm 1} \cdot s + b_{\rm 2}\cdot s^{2}~, 
\label{eq.ADparapoly}
\end{eqnarray}
and the height dependence is taken into account by the expression
\begin{eqnarray}
\theta_{\rm c}(h)  &=& \alpha \cdot E_{\rm thr}^{-\beta} ~~~,{\rm ~with}~E_{\rm thr}~{\rm in~MeV}
\label{eq.ADparaexp}
\\
\theta_{\rm cc}(h) &=& \gamma \cdot \theta_{\rm c}(h)~~~, {\rm ~with}~\gamma = \alpha^{\rm '}+ \beta^{\rm '}\cdot s~.
\label{eq.ADparaexp_cc}
\end{eqnarray}
The CORSIKA spectra are described properly using the 
following parameters

\begin{eqnarray}
(a_{\rm 0}, a_{\rm 1}, a_{\rm 2})  &=& (4.2489\cdot 10^{-1},  5.8371\cdot 10^{-1},  -8.2373\cdot 10^{-2}) \nonumber \\
(b_{\rm 0}, b_{\rm 1}, b_{\rm 2})  &=& (5.5108\cdot 10^{-2}, -9.5587\cdot 10^{-2}, 5.6952\cdot 10^{-2}) \nonumber \\
(\alpha, \beta)  &=& (0.62694,0.60590)  \nonumber \\
(\alpha^{\rm '}, \beta^{\rm '})  &=& (10.509,-4.9644)~. 
\label{param}
\end{eqnarray}

%
%
\subsection{Parameterisation of mean energy deposit}
The mean ionisation loss rate $\alpha_{\rm eff}$ in shower age $s$ can be approximated by
\begin{equation}
\label{eq.alpha_para_app}
\alpha_{\rm eff}~(s) ~=~\frac{c1}{(c_2 + s)^{c_3}} + c_4 + c_5 \cdot s~,~~~{\rm for}~E_{\rm cut}=1\,{\rm
MeV}~,
\end{equation}
with $c_1 = 3.90883,~ c_2 = 1.05301,~ c_3 = 9.91717, c_4 = 2.41715, c_5 = 0.13180$.

\end{document}